\documentclass[%
 apl,
 amsmath,amssymb,
 reprint,%
]{revtex4-1}

\usepackage{upgreek}
\usepackage{graphicx}
\usepackage{dcolumn}
\usepackage{bm}

\usepackage[utf8]{inputenc}
\usepackage[T1]{fontenc}
\usepackage{mathptmx}
\usepackage{etoolbox}

\makeatletter
\def\@email#1#2{%
 \endgroup
 \patchcmd{\titleblock@produce}
  {\frontmatter@RRAPformat}
  {\frontmatter@RRAPformat{\produce@RRAP{*#1\href{mailto:#2}{#2}}}\frontmatter@RRAPformat}
  {}{}
}%
\makeatother
\begin{document}

\preprint{AIP/123-QED}

\title{Dynamic response and roughening of ferroelectric domain walls driven at planar electrode edges} 

\author{Guillaume Rapin}
\altaffiliation[ ]{These authors contributed equally to this work, guillaume.rapin@unige.ch}
\affiliation{Department of Quantum Matter Physics, University of Geneva, 1211, Geneva, Switzerland}

\author{Sophia Ehrensperger}
\altaffiliation[ ]{These authors contributed equally to this work}
\altaffiliation[Now at ]{DACM, State of Geneva, Switzerland}
\affiliation{Department of Quantum Matter Physics, University of Geneva, 1211, Geneva, Switzerland}

\author{Cédric Blaser}
\altaffiliation[Now at ]{Federal Institute of Metrology METAS, Switzerland}
\affiliation{Department of Quantum Matter Physics, University of Geneva, 1211, Geneva, Switzerland}

\author{Nirvana Caballero}
\altaffiliation[]{nirvana.caballero@unige.ch}
\affiliation{Department of Quantum Matter Physics, University of Geneva, 1211, Geneva, Switzerland}

\author{Patrycja Paruch}
\altaffiliation[]{patrycja.paruch@unige.ch}
\affiliation{Department of Quantum Matter Physics, University of Geneva, 1211, Geneva, Switzerland}


\begin{abstract}
Understanding and controlling the motion, stability, and equilibrium configuration of ferroelectric domain walls is key for their integration into potential nanoelectronics applications, such as ferroelectric racetrack memories. Using piezoresponse force microscopy we analyse the growth and roughness of ferroelectric domains in epitaxial thin film Pb(Zr$_{0.2}$Ti$_{0.8}$)O$_3$, driven by the electric fields at straight edges of planar electrodes at two different temperatures. This device relevant geometry allows us to confirm that the domain walls are well described as 1-dimensional monoaffine elastic interfaces driven in random-bond disorder. However, we observe a progressive increase of roughness as initially flat domain walls move through the disorder landscape, which could prove a significant limiting factor for racetrack-type memories using ferroelectrics.
\end{abstract}

\maketitle

Ferroelectric materials, in which thin domain walls separate electrically reconfigurable domains with different polarisation orientation, are widely used in sensing, actuation, nonlinear optics, and information storage applications \cite{muralt_actuator,dawber_rmp_05_thin_films_ferroelectrics,qi_review,seungbum_book}. Recent research has focused on emergent structural and functional properties of these domain walls \cite{cherifi_natcomm_17_nonIsingDWs,meier_jp_15_functionalDW,seidel_natmat_09_BFO,guyonnet_am_11_DW_conduction} and their potential integration as active nanoelectronic device components \cite{salje_cpc_10_multiferroic_boundaries,catalan_RMP_12_DW_review,sharma_materials_19_functional}. For all such present and future applications, a detailed understanding of the fundamental physics underlying domain nucleation, growth, stability, and equilibrium configuration is essential. 

In this context, a useful approach allowing a predictive description of the motion and geometric properties of domain walls is to model them as elastic interfaces in a disordered medium \cite{giamarchi_domainwall_review,paruch_cras_13_DW_review}. This general statistical physics framework allows common features in systems as diverse as flux lines in type II superconductors \cite{blatter_rmp_94_vortex_review}, propagating fractures \cite{santucci_pre_07_fracture_statistics}, and proliferating cell fronts \cite{huergo_pre_2010_morphologycellfronts,chepizhko_pnas_2016_avalanchescellfronts,rapin_Sci_Rep_2021} to be compared. The behaviour of such systems is governed by the competition between elasticity, which tends to flatten the interface, and fluctuations in the potential energy landscape resulting from the disorder, which allow pinning. This competition leads to a self-affine equilibrium roughness configuration, whose characteristic power-law scaling is related to the dimensionality of the system and the type of disorder. When driven by an external force, elastic disordered systems present a complex and highly nonlinear dynamic response, with a thermally activated subcritical creep regime, and depinning when sufficiently high forces are applied.

In ferroelectrics, fundamental studies of roughness, memory effects, creep and depinning dynamics, and their characteristic avalanche statistics \cite{paruch_prl_05_dw_roughness_FE,paruch_prb_12_quench,tybell_prl_02_creep,jo_PRL_09_dw,Casals2020,tuckmantel_prl_2021_local} have led to the implementation of a wide range of prototype nanoelectronic devices based on domain walls, focusing in particular on racetrack-type memory applications \cite{mcgilly_natnanotech_15_DW_pinning, mcgilly_electrode_mod,mcmillen_apl_10_notches,mcquaid_nl_10_antinotches}. However, while in most device geometries the domain walls are driven by an electric field applied in a planar capacitor configuration \cite{hong_JAP_99_switching_dynamics,gruverman_prl_08_switching,McQuaid_NatComm_2017_injection,McConville_AdvFuncMat_2020_memristors}, in many fundamental studies the dynamic response is both induced and imaged using scanning probe microscopy (SPM). In such studies, the extremely local application of a highly inhomogeneous electric field \cite{blaser_njp_15_subcritical_switching} can provoke significant electrochemical effects \cite{kalinin_nano_11_electrochemical_SPM}, strongly dependent on both the polarisation orientation and switching history of the ferroelectric \cite{domingo_nanoscale_2019_surface}.  For comparison with theoretical models of a straight linear interface in equilibrium with the native disorder present in the sample, and for more direct assessment of what happens in device-relevant geometries, a useful alternative would be to exploit the nanoscale resolution of SPM for imaging, but to apply the external field using the straight edges of patterned macroscale electrodes. Such a protocol would establish a well defined initial domain wall position, determined by the electrode geometry, and allow the domain walls to subsequently move away from the electrode edge and into the pristine region beside it, reflective of the disorder potential landscape established during sample growth and unperturbed by electrochemical effects of domain writing using a biased tip. While pairs of surface electrodes can be used to establish an electric field oriented in the film plane to drive the switching of in-plane polarisation components \cite{Sharma_AdvFuncMat_2019_conformational,Guy_AdvMat_2021_anomalous}, switching out-of-plane polarisation components requires an electric field perpendicular to the film plane.

\begin{figure*}[t!]
    \centering
    \includegraphics[width=\linewidth]{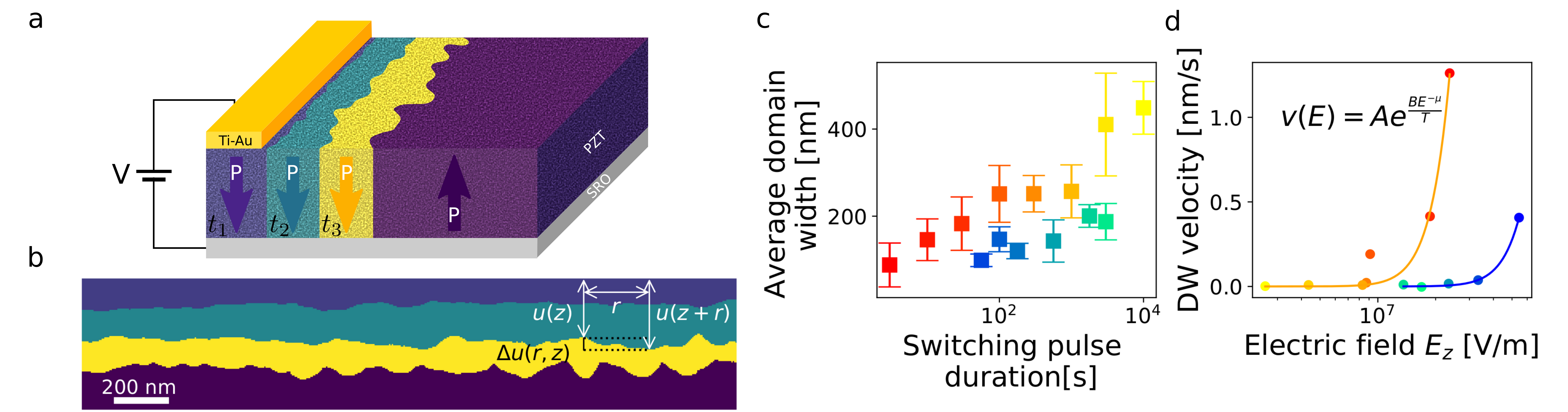}
    \caption{\emph{Dynamics of domain walls at electrode edges} (a) Schematic representation of the measurement geometry, with the outward growth of a down-polarised stripe domain (blue, teal, yellow), driven by the out-of-plane component of the fringing electric fields between the surface Ti-Au electrodes and the SrRuO$_3$ bottom electrode into the pristine up-polarised as-grown region of the sample (aubergine) (b) Composite illustration of the displacement $u(z)$ of the 180° domain wall with respect to the electrode edge, obtained from PFM phase images taken after the application of 10 V writing pulses for 3 (blue), 30 (teal), and 300 s (yellow) to Au/Ti electrodes patterned on a Pb(Zr$_{0.2}$Ti$_{0.8}$)O$_3$ thin film at 100$^\circ$ C. Initially relatively flat, the down-polarised domains grow outwards from the electrode edge into the up-polarised as-grown regions of the film (aubergine) with increasing domain wall roughness, which can be quantified via the correlation of the relative displacements $\Delta u(r, z) = u(z) - u(z+r)$ measured between pairs of points separated by a distance $r$. (c) The average width of the growing domains increases logarithmically with increased switching pulse duration. (d) Domain wall velocity as a function of the out-of-plane component of the electric field at the edge of the electrode, showing a highly nonlinear dependence consistent with a creep process $v(E)=e^{\frac{-U_c}{k_BT}(\frac{E_c}{E})^\mu}$. Comparison between room temperature (23$^\circ$ C, blue--green colour scale) and high temperature (100$^\circ$ C, red--yellow colour scale) measurements shows the significant effects of thermal activation characteristic of creep dynamics. We use the same colour scale in (c) and (d) to show the correspondence between points in both plots.}
  \label{domain_growth}
\end{figure*}

Here, we report on such a study, following domain wall dynamics and roughness of initially flat domain wall segments in Pb(Zr$_{0.2}$Ti$_{0.8}$)O$_3$ thin films, driven by the fringing electric fields at the edges of straight patterned electrodes, as schematically illustrated in Fig.~\ref{domain_growth}(a). We find that the domain walls are well described as monoaffine 1-dimensional elastic interfaces in random bond disorder, with measurements at both 23 and 100$^\circ$ C demonstrating the importance of thermal activation in the creep regime, and allowing us to extract the value of the creep exponent $\mu = 0.21 \pm 0.02$ with very high precision. We observe a progressive roughening of the domain walls as they move away from the electrode edges, with the roughness exponent increasing from $\zeta = 0.5--0.6$ to $\zeta = 0.7--0.8$, suggesting a qualitative difference between the disorder potential landscape of the as-grown sample, and after high electric field application This picture is compatible with observations in numerical simulations of interfaces in a Ginzburg-Landau model~\cite{caballero_JSTAT_2021_ac}. The observed roughening of the domain walls, which increases their effective width, may have significant consequences for the use of ferroelectrics in racetrack memory applications.

All measurements were performed on a $\sim$270 nm Pb(Zr$_{0.2}$Ti$_{0.8}$)O$_3$ thin film, epitaxially grown with a \mbox{35 nm SrRuO$_3$} back electrode on (001) oriented single crystal SrTiO$_3$ by off-axis RF magnetron sputtering \cite{blaser_apl_12_CNT_FE}. As detailed in Supplementary Materials, the film shows high crystalline and surface quality, and presents an out-of-plane up-oriented monodomain polarisation. 50 nm Au/5 nm Ti top electrodes with extended straight-edged sections were deposited by e-beam evaporation after photolithographic patterning. 10 V pulses of varying duration were applied to selected electrodes using ZN50R-10-BeCu needle contacts in a \textit{Lakeshore} cryogenic probe station to induce polarisation switching under the electrode in a standard parallel plate capacitor geometry, followed by the outward growth of the down-polarised domains at the electrode edges, driven by the exponentially decreasing out-of-plane component of the capacitor fringing field. To examine the effects of thermal activation, this polarisation switching was carried out at 23$^\circ$ C and 100$^\circ$ C. The resulting domain configuration was imaged by piezoresponse force microscopy (PFM) in ambient conditions and analysed with the \textit{Hystorian} materials science data analysis package \cite{musy_ultramicroscopy_2021_hystorian} to extract the position and follow the evolution of the 180° domain walls.

For short switching pulses, we initially observe the presence of small discontinuous down-polarised regions along the electrode edge, which coalesce into a continuous and progressively wider stripe domains when longer writing pulses are applied (see Supplementary Materials for PFM phase data) and the domain walls move further away from the electrode into the up-polarised bulk of the sample. From the images of the growing domains, we extract the transverse displacement $u(z,t)$ along the longitudinal coordinate $z$ of the domain walls with respect to the electrode edge for a given switching pulse duration $t$, as shown in Fig.~\ref{domain_growth}(b).  The average domain width $\langle u(z,t) \rangle$, obtained by averaging $u(z,t)$ across a minimum of 10 separate images from at least 4 independent electrodes, increases logarithmically with the switching pulse duration, as can be seen in Fig.~\ref{domain_growth}(c). While the process is qualitatively similar, we observe significantly earlier onset and faster rates of domain growth at the higher temperature. From these measurements, we extract the domain wall velocity as the increase of the average domain width $\langle u(z,t_2) \rangle - \langle u(z,t_1) \rangle$ over the interval between two sequential writing times $t_2 - t_1$, and correlate it with the out-of-plane electric field $E_z$ at the edge of the electrode, obtained via Comsol modelling (as detailed in Supplementary Materials). We find a highly nonlinear dynamic response, which can be very well described as a creep process - the very slow, thermally activated dynamics for a subcritical driving force, characterised by jerky, stochastic jumps of an interface between different local minima in a highly heterogeneous potential energy landscape:
\begin{equation}
    v(E)=e^{\frac{-U_c}{k_BT}(\frac{E_c}{E})^\mu}
    \label{creep_equation}
\end{equation}
where $U_c$ is the characteristic energy barrier, $T$ the temperature, $k_B$ the Boltzmann constant, $E$ the applied electric field driving domain wall motion, and $E_c$ the critical field for depinning. Importantly, since fitting is performed self-consistently on the full dataset for both temperatures, a far more precise estimate of the dynamic exponent $\mu = 0.21 \pm 0.02$ can be obtained (as detailed in Supplementary Materials) than in previous studies \cite{tybell_prl_02_creep,paruch_prl_05_dw_roughness_FE}.

\begin{figure*}
    \centering
    \includegraphics[width=\linewidth]{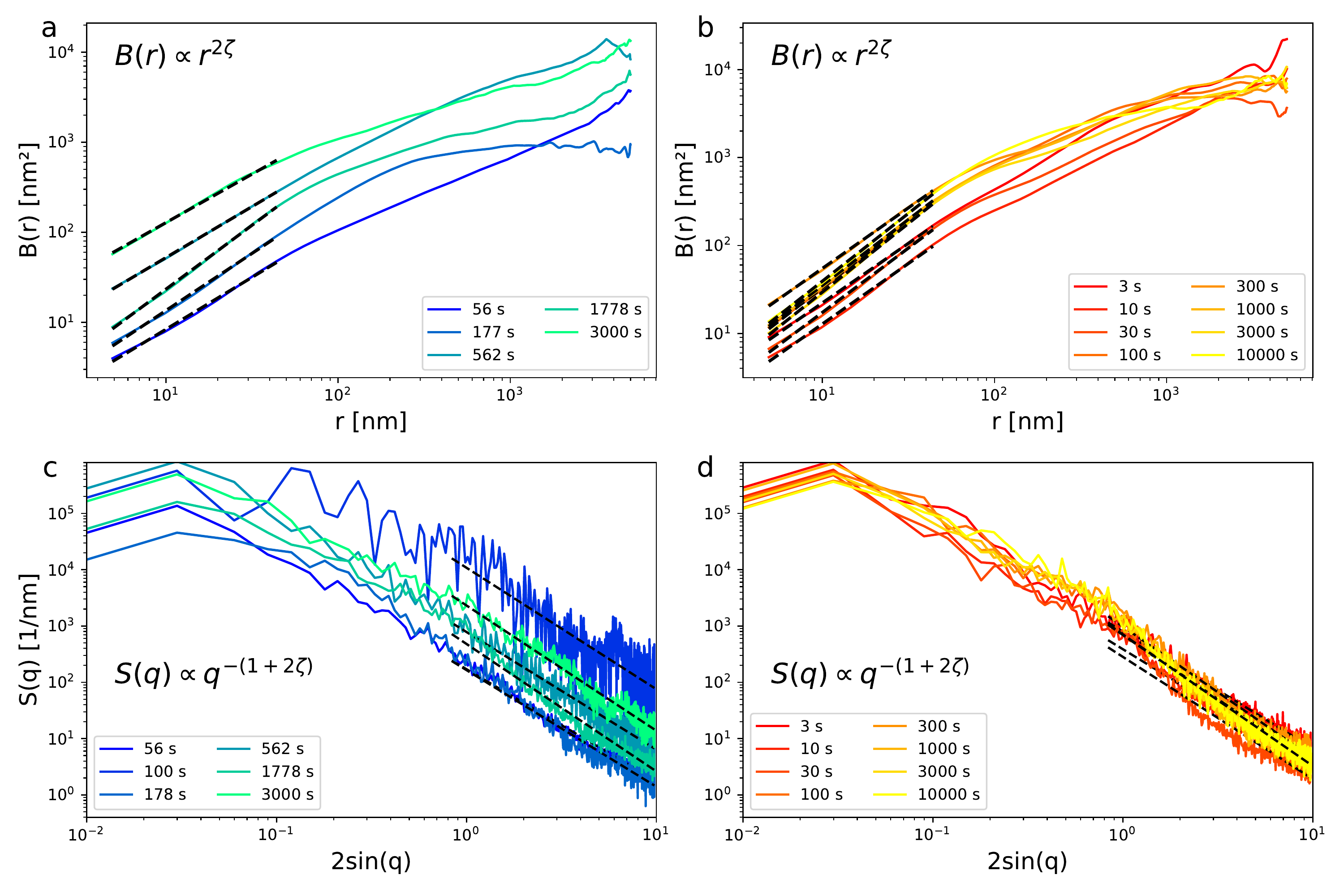}
    \caption{\emph{Evolution of domain wall roughness during domain growth} Averaged roughness correlation function $B(r) = \left<|\Delta u(r,z)|^2\right>$ of domain walls written with pulse times ranging between 56 and 3000 s for 23$^\circ$ C (a), and between 3 and 10000 s for 100$^\circ$ C (b), showing power law scaling $B(r) \sim r^{2\zeta}$ at short lengthscales $r$. Averaged structure factor $S(q) = \overline{\langle (\tilde{u}(q,t)\tilde{u}(-q,t))^n\rangle}$ of domain walls written with pulse times ranging between 56 and 3000 s for 23$^\circ$ C (c), and between 3 and 10000 s for 100$^\circ$ C (d), showing power law scaling $S(q) \propto q^{-(1+2\zeta)}$ at large reciprocal lengthscales $q$. Fitting (black dashed lines) allows the characteristic roughness exponent $\zeta$ to be extracted from both real space and reciprocal space analysis.} 
    \label{Br_Sq}
\end{figure*}

During the growth of the stripe domain, we also observe a pronounced evolution of their geometric configuration.  As shown in Fig.~\ref{domain_growth}(b), initially relatively flat domain walls near the electrode edges roughen visibly as they move progressively further away. This roughening can be quantified by examining the relative displacements $\Delta u(r, z)=u(z)-u(z+r)$ between two points along the wall separated by a distance $r$. Extracting the correlation function of these relative displacements 
\begin{equation}
B(r,t) = \overline{\left<|\Delta u(r,t)|^2\right>}\sim r^{2\zeta}
\label{equ:Br}
\end{equation}
where $\left<...\right>$ and $\overline{...}$ signify the average over different $z$ values for a single domain wall segment, and the average over different realizations of disorder, respectively. We find overall a similar level of roughness for the domain walls driven at 23$^\circ$ C (Fig. \ref{Br_Sq}(a)) and 100$^\circ$ C (Fig. \ref{Br_Sq}(b)), comparable to previous reports of roughening at SPM-tip-patterned ferroelectric domain walls in PZT films subjected to thermal heat--quench cycles \cite{paruch_prb_12_quench}. At both 23$^\circ$ C and 100$^\circ$ C, we find that the domain walls driven by the shortest switching pulses show the expected power-law scaling growth of $B(r)$ only at short lengthscales $r$ up to 40--50 nm, followed by an apparent saturation reflecting their essentially flat nature at large lengthscales and initial times \cite{kolton_prl_05_flat_interface,caballero_prb_2020_qEW-GL}. For longer writing times $B(r)$ increases and, particularly at the higher temperature, the power-law scaling region appears to extend somewhat further to over 200 nm.  This behaviour indicates that the electrode edge does not have a significant effect on the domain wall roughness exponents.

To further confirm these observations, and compare the scaling exponents extracted with both approaches, we also carried out a reciprocal space analysis of the roughness, extracting the structure factor
\begin{equation}
    S(q,t) = \overline{\langle (\tilde{u}(q,t)\tilde{u}(-q,t))^n\rangle},
    \label{equ:Sq}
\end{equation}
where $\tilde{u}(q,t)$ is the Fourier transform of the displacement field which defines the domain wall position
\begin{equation}
    \tilde{u}(q,t) = \frac{1}{L}\int{dz\, u(z,t)e^{-iqz}}.
    \label{equ:uq}
\end{equation}
Although $B(r,t)$ and $S(q,t)$ are related, containing the same geometric information about the domain wall roughness
\begin{equation}
    B(r,t) = \int{\frac{dz}{\pi}[1-\cos(qr)]S(q,t)},
    \label{equ:BS}
\end{equation}
the latter provides more reliable estimates of the roughness exponent value when sufficient statistics are available \cite{bustingorry_jpcm_2021_numerical}. As shown in Fig. \ref{Br_Sq}(d,e), we observe the expected power-law scaling of the structure factor $S(q) \propto q^{-(1+2\zeta)}$ at high reciprocal lengthscales $q$ for the domain walls driven at 23$^\circ$ C and 100$^\circ$ C, respectively.

\begin{figure*}
    \centering
    \includegraphics[width=\linewidth]{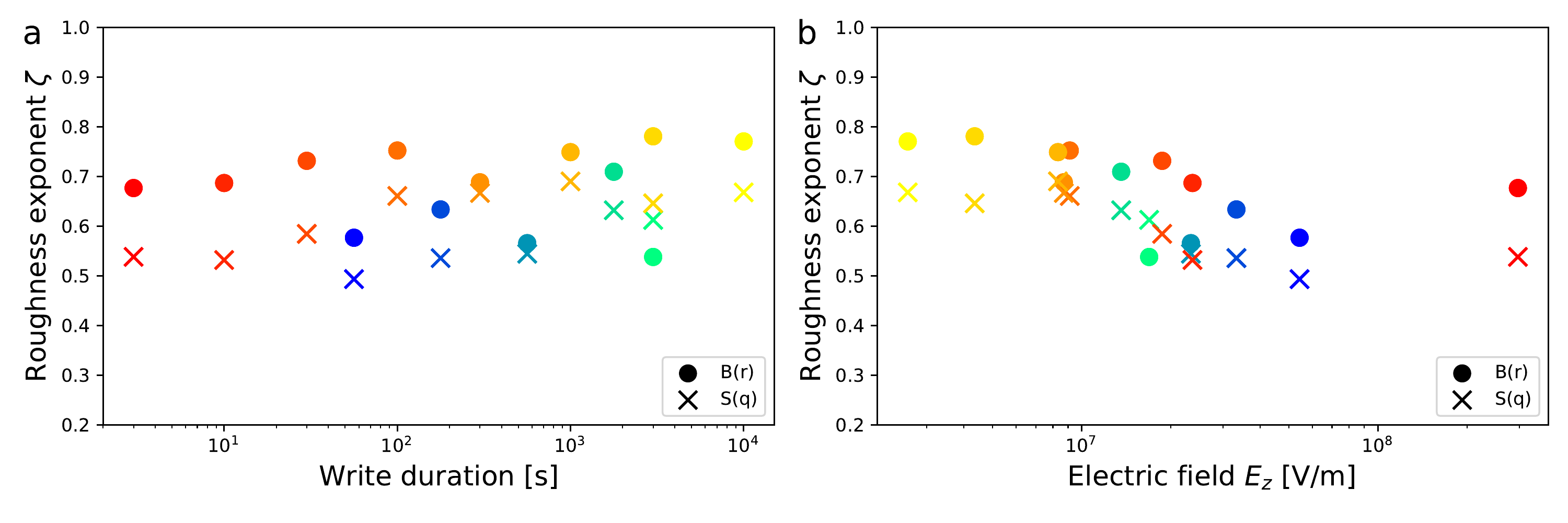}
    \caption{\emph{Evolution of the domain wall roughness exponent during domain growth} $\zeta$ values extracted from fitting power-law scaling of $B(r)$ (circles) and $S(q)$ (crosses), and followed as a function of (a) the switching pulse duration and (b) the out-of-plane electric field. While at room temperature (23$^\circ$ C, blue-green colour scale), $\zeta$ values cluster uniformly around 0.5--0.6, at high temperature (100$^\circ$ C, red-yellow colour scale) $\zeta$ values increase to 0.7--0.8 for long switching pulse duration, as the domain walls move further away from the electrode edges into a region of lower electric field intensity. The colour scales are used to help the reader to assess the correspondence between points in (a) and (b).}
    \label{zetas_br_sq}
\end{figure*}

The roughness exponent $\zeta$, whose value depends on the dimensionality of the system, the nature of the disorder potential, and the range of the elastic interactions, can then be obtained by fitting this characteristic power-law scaling of $B(r)$ and $S(q)$. We find that the roughness exponents obtained from both the real space and reciprocal space analysis are in excellent agreement, although the latter consistently gives slightly lower values. As can be seen in Fig. \ref{zetas_br_sq}(a), at room temperature $\zeta$ values cluster uniformly around 0.5--0.6, but increase to 0.7--0.8 for long switching pulse duration at the higher temperature. 

Multiscaling analysis of the probability distribution function of relative displacements following \cite{santucci_pre_07_fracture_statistics,guyonnet_prl_12_multiscaling} (detailed in Supplementary Materials) confirms that the observed scaling appears to be monoaffine, as expected for elastic interfaces subject to weak collective pinning by random disorder \cite{halpin-healy_pra_91_directed_polymers,mezard_jdpi_91_replica,rosso_jstatmech_05_gaussian}, rather than multiaffine \cite{nattermann_prb_90_creep_domainwall,barabasi_pra_92_multifractiality,kolton_prl_05_flat_interface}.

These observations of ferroelectric domain wall roughness and dynamics in our specific measurement geometry reveal a number of important features. First, by comparing domain walls driven at 23$^\circ$ C and 100$^\circ$ C we explicitly confirm the very significant role of thermal activation during creep motion, with speeds up to 2 orders of magnitude higher observed for comparable electric fields at the higher temperatures, an effect that is particularly important at low fields. This thermal activation allows a much larger stripe domain to grow for a given switching pulse duration, in spite of the exponentially decreasing magnitude of the out-of-plane component of the electric field as one moves further away from the electrode edge. Second, the very different aspect presented by the domain walls in the immediate vicinity of and furthest away from the electrode edges, with over twofold increase in roughness $B(r)$ and an evolution of the roughness exponent $\zeta$ from $\sim 0.5$ to $\sim 0.8$ points towards qualitative differences in the disorder potential landscape in the two cases. 

Indeed, past studies of the effects of high intensity electric fields applied via an SPM tip demonstrated that polarisation reversal can be accompanied by the injection and reorganisation of charged defects, long-lasting electrochemical changes on the ferroelectric surface, and even significant material degradation for the highest field intensity \cite{kalinin_nano_11_electrochemical_SPM,ievlev_acsami16_tip_induced_electrochemistry,domingo_nanoscale_2019_surface}. The resulting patterned domains commonly show marked pinning of domain walls at their initial position, even after subsequent field application or heating \cite{tuckmantel_prl_2021_local,paruch_prb_12_quench}. We believe the strong electric field right under the electrode and at its edge, comparable to that applied under a biased SPM tip \cite{blaser_apl_12_CNT_FE,blaser_njp_15_subcritical_switching}, has similar effects in our sample. 

We note, moreover, that the relatively flat geometry of the domain walls in the immediate vicinity of the electrode edges, in spite of the higher electric field intensity, suggests the system is not simply in a depinning regime, where in fact higher growth of domain wall roughness may be expected \cite{lopez_superroughening,purrello_superroughening}. However, as the domain walls progressively move away from the electrodes into the region of decreasing electric field intensity, their growing roughness, increasing roughness exponent values, and the extension of the power-law scaling region to higher $r$ suggest that here the system is more effectively relaxing toward equilibrium with the pristine disorder potential landscape of the as-grown sample. 

Values of the roughness exponent $\zeta$ in the 0.6--0.7 range had previously been observed in for both artificial and naturally occurring domain walls in Pb(Zr$_{0.2}$Ti$_{0.8}$)O$_3$ \cite{pertsev_jap_11_ceramics,guyonnet_prl_12_multiscaling}, and together with dynamic exponent $\mu$ of 0.21 would be compatible with the theoretical predictions for 1-dimensional elastic interfaces in random bond disorder \cite{nattermann_epl_87_rfield_rbond,agoritsas_physb_12_DES}. The random bond disorder at the origin of the observed domain wall roughening can be interpreted as site-to-site fluctuations in the energy barrier that the polarisation has to overcome during 180° switching~\cite{caballero_prb_2020_qEW-GL}. These fluctuations can be related to uncharged defects in the crystalline order or locally varying surface morphology~\cite{paruch_cras_13_DW_review,lu2019ferroelectric}.

Our observations of progressively increasing roughness as initially flat domain walls move through the disorder landscape are particularly pertinent for potential ferroelectric-based racetrack memory applications \cite{mcgilly_natnanotech_15_DW_pinning,mcgilly_electrode_mod,mcmillen_apl_10_notches,mcquaid_nl_10_antinotches}, where the resulting broadening of the information-carrying region could prove a significant limiting factor. A possible solution would be to focus on ferroelastic ferroelectric domain walls \cite{Scott_superdomain_dynamics,salje_ferroelastic_DW_devices,Barzilay_BTO_DWs}, where additional strain terms could help maintain straighter walls.

\section*{Author contributions statement}
P.P. designed and supervised the work. S.E. performed the PFM switching measurements, and carried out preliminary analysis with assistance from C.B. G.R carried out the full multiscaling data analysis. G.R., N.C., and P.P. wrote the manuscript. All authors contributed to the scientific discussion and manuscript revisions. 

The authors declare no competing interests.

\begin{acknowledgments}
This work was partially supported by  Division II of the Swiss National Science Foundation under project 200021\_178782. N.C. acknowledges support from the Federal Commission for Scholarships for Foreign Students for the Swiss Government Excellence Scholarship (ESKAS No. 2018.0636). 
\end{acknowledgments}

\section*{Data Availability Statement}
The data that support the findings of this study are openly available in \textit{Yareta} at http://doi.org/10.26037/yareta:pmummlrfbneihiwchlsejykinu.

%


\appendix
\clearpage

\onecolumngrid
\begin{center}
\medskip
\begin{large}
\textbf{Supplementary material for \\ ``Dynamic response and roughening of ferroelectric domain walls driven at planar electrode edges''}
\end{large}
\end{center}

\bigskip

\twocolumngrid

\subsection{PFM measurements of domain growth at electrode edges}
\begin{figure*}[h]
    \centering
   \includegraphics[width=\linewidth]{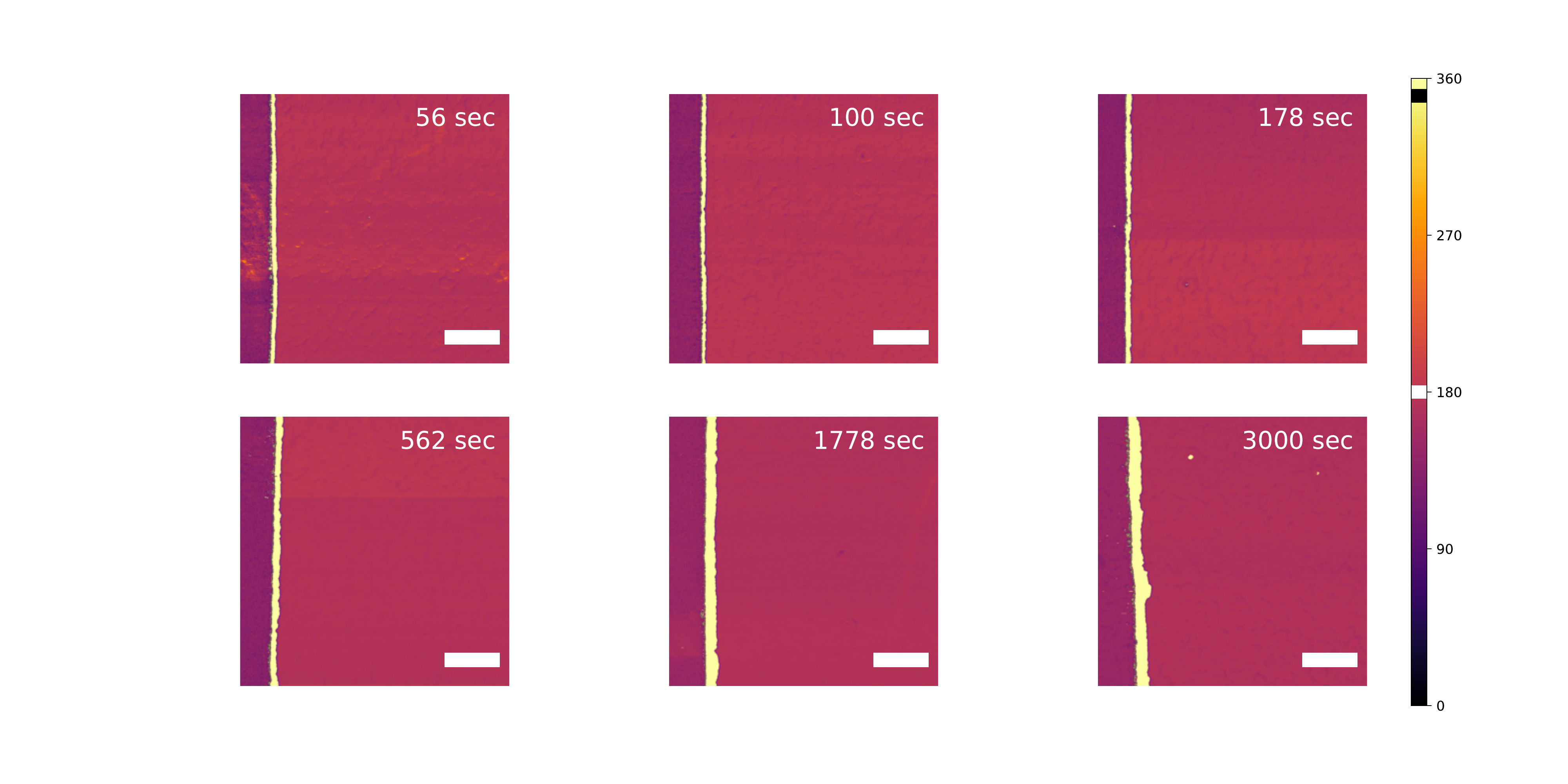}
    \caption{\emph{PFM phase images of domain growth at electrode edges at 23$^\circ$C.}  Domains written with 10 V pulses applied to the top electrode for the indicated switching pulse duration. The white bar represents 1 $\upmu$m, and all images are shown at the same PFM phase scale, with the white and black regions on the scale representing the approximate value of the down and up polarised domains, respectively.}
    \label{fig:PFM_supp_23_phase}
\end{figure*}

\begin{figure*}[h]
    \centering
   \includegraphics[width=\linewidth]{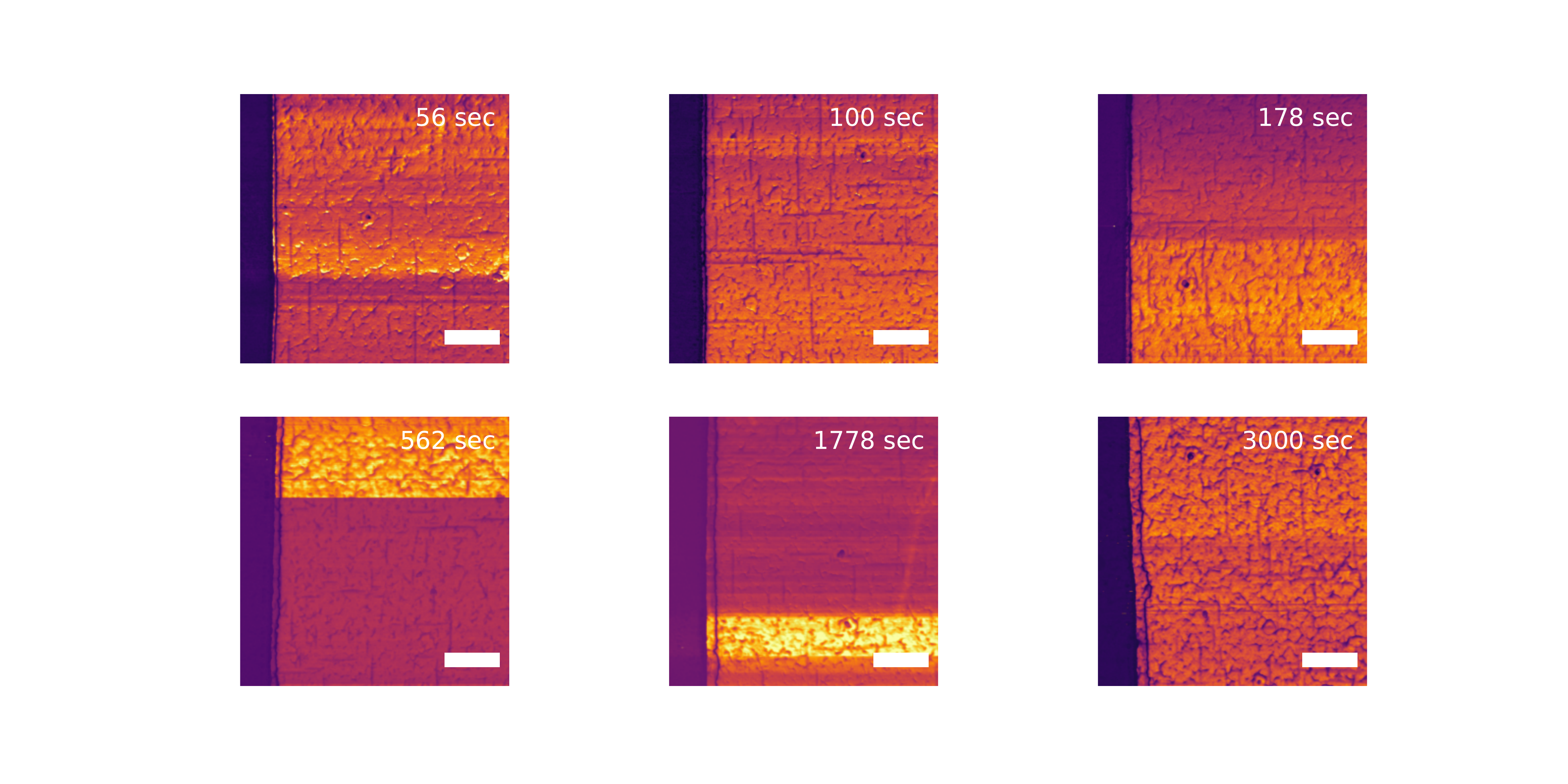}
    \caption{\emph{PFM amplitude images of domain growth at electrode edges at 23$^\circ$ C.}  Acquired concurrently with the measurements shown in Fig. \ref{fig:PFM_supp_23_phase}, the amplitude images recapitulate the same information, with the 180° domain walls visible as a narrow dark line to the right of the electrode, itself corresponding to a region of minimum amplitude at the left of each image, since it blocks our PFM signal. The white bar represents 1 $\upmu$m.}
    \label{fig:PFM_supp_23_amplitude}
\end{figure*}

\begin{figure*}[h]
    \centering
   \includegraphics[width=\linewidth]{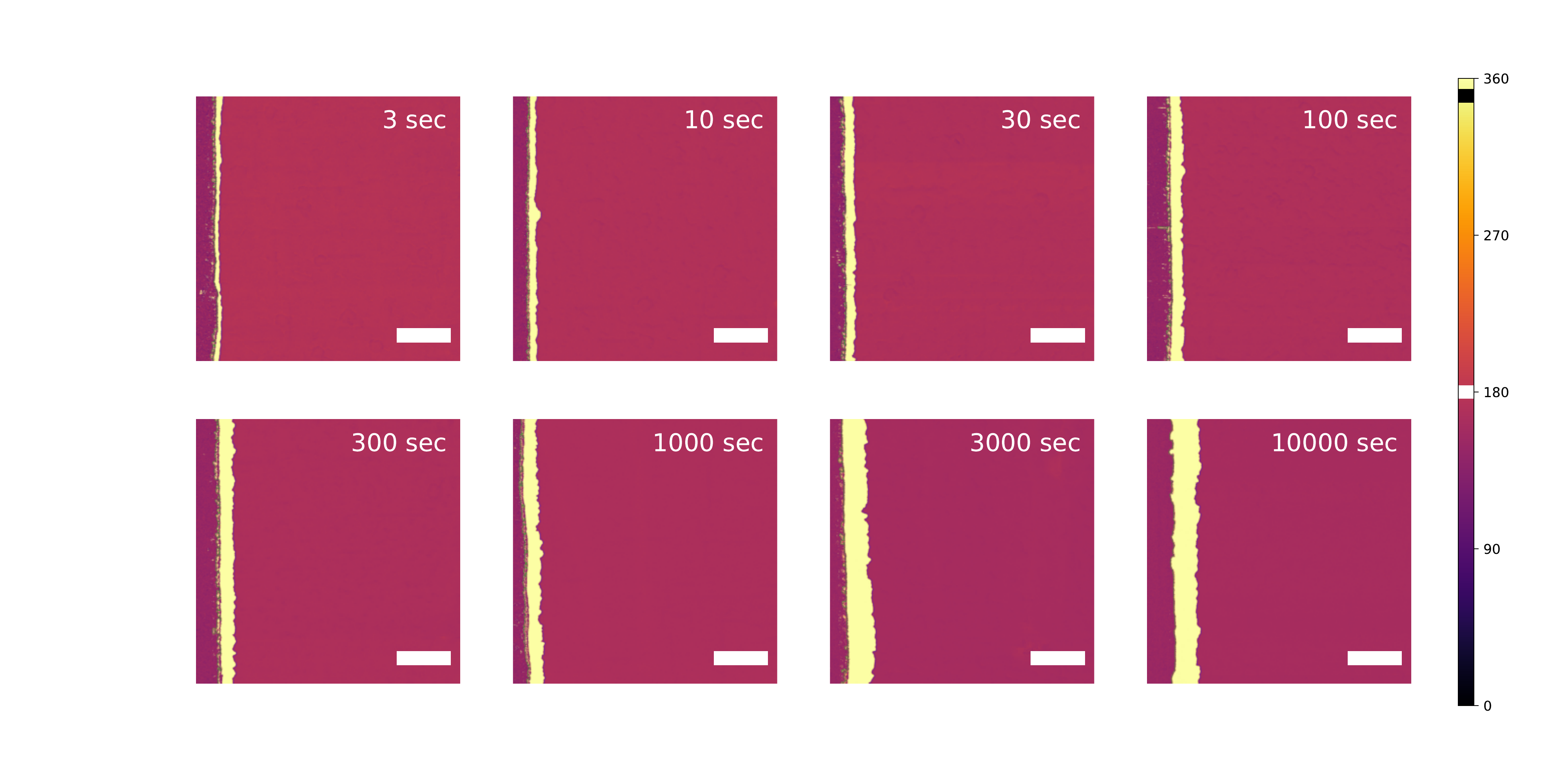}
    \caption{\emph{PFM phase images of domain growth at electrode edges at 100$^\circ$ C.}  Domains written with 10 V pulses applied to the top electrode for the indicated switching pulse duration. The white bar represents 1 $\upmu$m, and all images are shown at the same PFM phase scale. with the white and black regions on the scale representing the approximate value of the down and up polarised domains, respectively.}
    \label{fig:PFM_supp_100_phase}
\end{figure*}

\begin{figure*}[h]
    \centering
   \includegraphics[width=\linewidth]{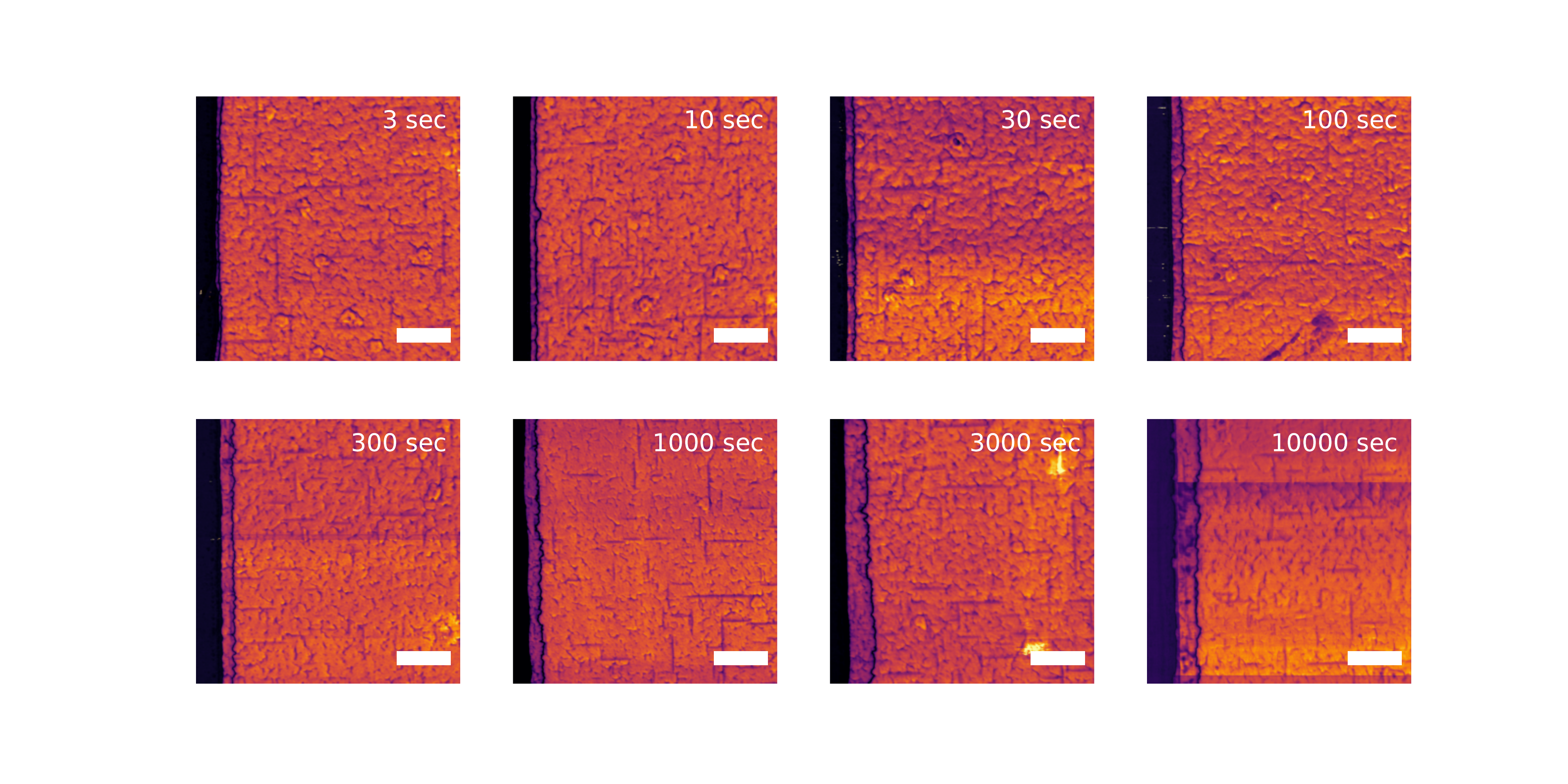}
    \caption{\emph{PFM amplitude images of domain growth at electrode edges at 100$^\circ$ C.}  Acquired concurrently with the measurements shown in Fig. \ref{fig:PFM_supp_100_phase}, the amplitude images recapitulate the same information, with the 180° domain walls visible as a narrow dark line to the right of the electrode, itself corresponding to a region of minimum amplitude at the left of each image, since it blocks our PFM signal. The white bar represents 1 $\upmu$m.}
    \label{fig:PFM_supp_100_amplitude}
\end{figure*}

To follow the growth of the down-polarised domains into the up-polarised as-grown regions of the sample, vertical piezoresponse force microscopy (PFM) imaging was carried out in a Bruker Dimension V AFM system at ambient conditions, using Bruker MESP tips, with a \mbox{20 kHz} drive frequency, \mbox{3000 mV} drive amplitdue, and \mbox{3 $\upmu$m/s} tip scanning velocity). After the application of a 10 V switching pulse for the chosen duration, concurrent PFM phase and amplitude 5 $\times$ 5 $\upmu$m\textsuperscript{2} images were taken along the electrode edge with a 1024 $\times$ 1024 pixel resolution, as can be seen in Figs. \ref{fig:PFM_supp_23_phase}-- \ref{fig:PFM_supp_100_amplitude} for domains switched at 23$^\circ$ C and 100{$^\circ$} C, respectively. Given the intrinsically almost perfectly binary nature of phase imaging (180° phase difference with relatively low noise) with a steep change between the two phase states corresponding to up- vs. down-polarised domains, convoluted essentially only by the tip size, vertical PFM phase provides the most reliable and least noisy way to precisely determine the domain wall position.  The amplitude channel can be used, tracking its minimum as the position of the domain wall in each scan line, and gives comparable domain wall geometry, but with far more noise. This is because the amplitude channel is more prone to variability via electrostatic and electrochemical effects, and as a result of contact resonance variations giving rise to crosstalk with the sample topography.

\subsection{Image processing
\label{Sup:Image_processing}}
Each of the 1024 $\times$ 1024 px vertical PFM phase images, with phase values normalised on a 0--360$^\circ$ range, were offset to obtain comparable minimum values (centered around $180^\circ$, crimson colour in Figs. \ref{fig:PFM_supp_23_phase},\ref{fig:PFM_supp_100_phase}) for the as-grown up-polarised domains, and maximum values (centered around $360^\circ$, yellow colour in Figs. \ref{fig:PFM_supp_23_phase},\ref{fig:PFM_supp_100_phase}) for the down-polarised domains growing outwards from the electrode edges. The images were then binarised using in-house developed algorithms within the Hystorian materials science data analysis Python package \cite{musy_ultramicroscopy_2021_hystorian}, allowing the positions of the electrode edges and domain walls to be identified, and their $x,y$ coordinates extracted. Finally, the relative front displacement $u(z)$ was obtained as the difference in the $x$ coordinates of the electrode edge and domain wall for each $y$ coordinate value.

\subsection{Modelling electric field intensity at electrode edges}

\begin{figure*}[h]
    \centering
    \includegraphics[width=\linewidth]{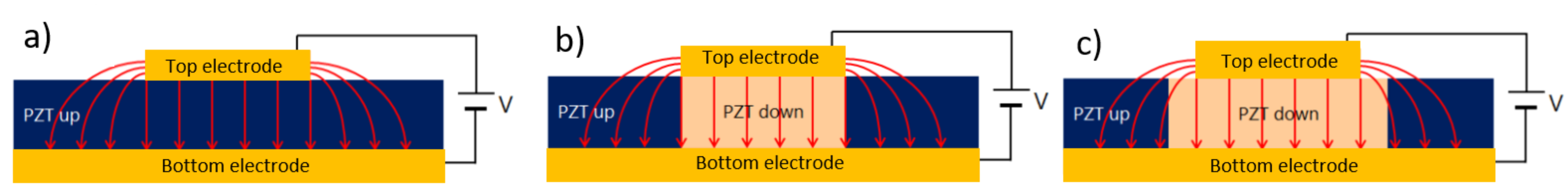}
    \caption{\emph{Schematic representation of polarisation reversal in a planar capacitor geometry}  (a) The initially up-polarised PZT thin film, with a SrRuO$_3$ epitaxial bottom electrode and Au/Ti patterned top electrodes, between which 10 V pulses are applied, giving rise to uniform, out-of-plane, high intensity electric field under the top electrode, and fringing fields extending from its edges (field lines represented in red). As a result, initial domain nucleation and growth occurs directly under the top electrode (b), followed by propagation of domain walls away from the electrode edges (c).}
    \label{fig:switching_schematic}
\end{figure*}

When voltage pulses are applied across a ferroelectric film sandwiched between macroscopic electrodes in a planar capacitor geometry, as schematically illustrated in Fig. \ref{fig:switching_schematic}(a--c), previous studies have shown that polarisation switching proceeds by the nucleation, growth, and coalescence of multiple domains \cite{hong_JAP_99_switching_dynamics,gruverman_prl_08_switching} in the intense, homogeneous and purely out-of-plane electric field within the capacitor. Once the ferroelectric volume within the capacitor is fully switched into the polarisation state orientated parallel to the applied electric field, much slower outward domain growth is possible, driven by the fringing fields extending beyond the edges of the electrodes. 

\begin{figure*}[h]
    \centering
    \includegraphics[width=\linewidth]{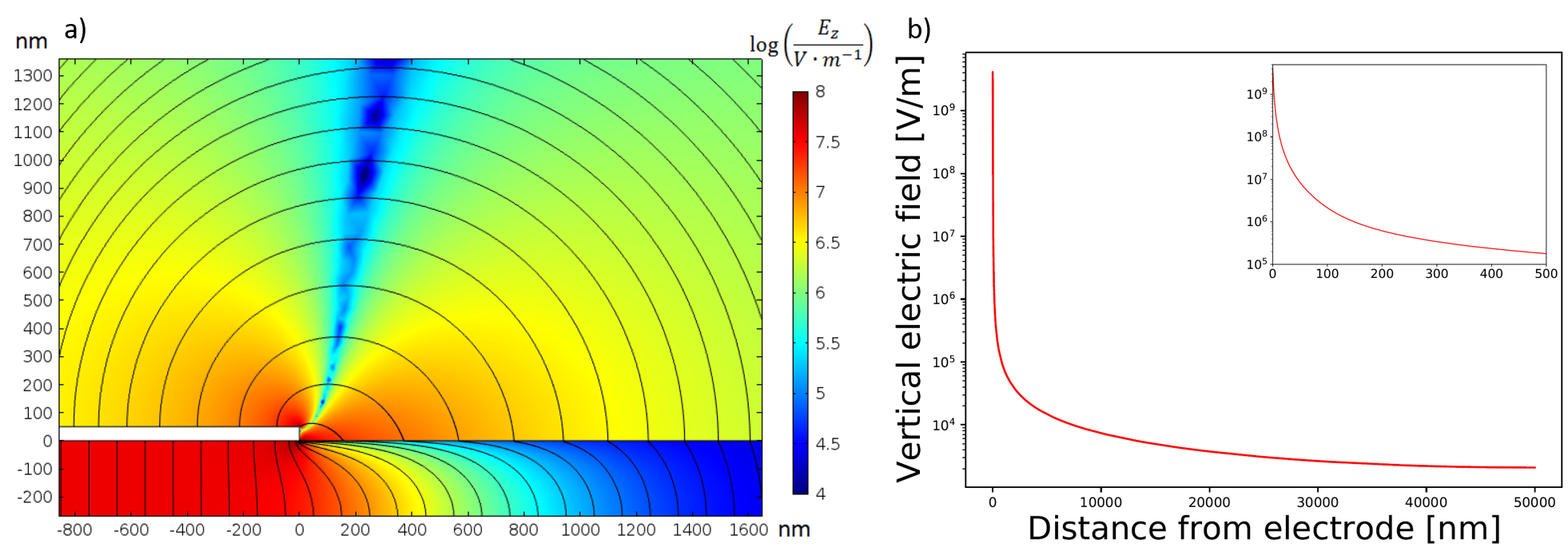}
    \caption{\emph{Electric field driving domain wall dynamics at electrode edges}  (a) Finite element simulation of the out-of-plane component of the electric field $E_z$ in the planar capacitor geometry of the measurements, and (b) the corresponding values of $E_z$ extracted at the ferroelectric surface, as a function of the distance from the electrode edge. The inset shows the $E_z$ values at length scales corresponding to the width of the domains imaged in our experiments.}
    \label{fig:electric_field}
\end{figure*}

Using Comsol finite element 2D (slab) simulations of a 270 nm thick, 10 $\upmu$m long PZT film with its lower boundary fixed at ground,  and a 55 nm thick, 5 $\upmu$m wide perfectly conducting top electrode to which 10 V potential could be applied, we numerically simulated the fringing electric field to extract its out-of-plane component, shown in Fig. \ref{fig:electric_field}(a). As can be seen in Fig. \ref{fig:electric_field}(b), the intensity of the fringing fields decreases very rapidly as a function of distance from the electrode edge.  We therefore expect progressively slower domain wall motion and less significant effects of surface and bulk charge dynamics as the domain walls are driven further from the electrodes.

\subsection{Extracting the creep exponent $\mu$}
The domain wall velocity during creep depends on both the temperature $T$ and the vertical component of the electric field $E_z$ which drives the motion. We therefore carried out a self-consistent two-dimensional surface fitting of the creep equation (Eq. 1 of the main text) to the cloud of points combining data obtained from switching at both 23$^\circ$ C and 100$^\circ$ C, as shown in Fig. \ref{fig:electric_field}(a). Seed values for the two-dimensional fitting parameters were obtained from fits carried out separately on the datasets obtained at the two different temperatures, shown for comparison in Fig. \ref{fig:electric_field}(b). Based on these preliminary values, the two-dimensional fits were processed using the Python \emph{Scipy} library and the \emph{curve\_fit} function.
\begin{figure*}[h]
    \centering
    \includegraphics[width=\linewidth]{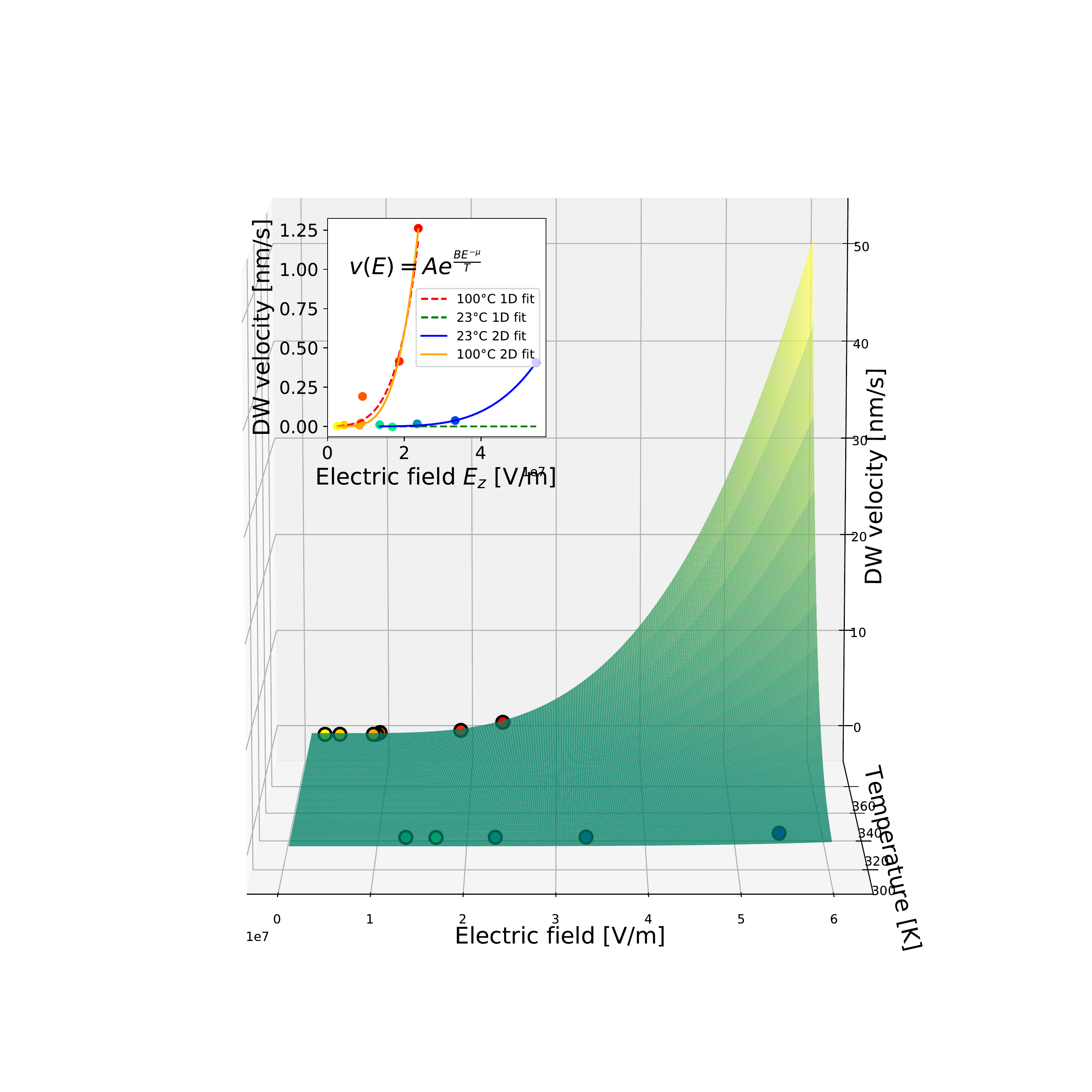}
    \caption{\emph{Fitting domain wall motion as a creep process $v(E)=e^{\frac{-U_c}{k_BT}(\frac{E_c}{E})^\mu}$ } Domain wall velocity as a function of the out-of-plane electric field and temperature, self-consistently fitting the data obtained at both 23$^\circ$ C and 100$^\circ$ C as a two-dimensional surface. The inset shows domain wall velocity as a function of the out-of-plane component of the electric field at the edge of the electrode, where solid lines are extracted from the two-dimensional fit, and the dashed lines from fits done separately on the data obtained at the two different temperatures. 23$^\circ$ C data are shown on the blue--green colour scale, and 100$^\circ$ C data on the red--yellow colour scale.}
    \label{fig:creep_fitting}
\end{figure*}

\subsection{Multiscaling analysis of the probability distribution function of relative displacements}

The statistical distribution of the fluctuation of the domain wall position can be quantified by defining the probability distribution function (PDF) of the relative displacements $\Delta u(r,z)$ at a given length scale $r$
\begin{equation}
    P[\Delta u(r,z)] = \frac{1}{N}\int{dz\cdot\Delta u(r,z)}
\end{equation}
where the factor $N$ ensures normalisation. The central moments of this PDF, reflecting its characteristic scaling properties \cite{agoritsas_physb_12_DES}, are the real-space displacement autocorrelation functions
\begin{equation}
\sigma_n(r)=\overline{\left<|\Delta u(r)|^n\right>}\sim r^{n\zeta_n},
\end{equation}
where $\zeta_n$ are the associated scaling exponents for the $n$th moment.

For monoaffine systems, such as 1-dimensional equilibrated interfaces at zero temperature in weak collective pinning, the PDF is well approximated by a Gaussian function  \cite{halpin-healy_pra_91_directed_polymers,mezard_jdpi_91_replica,rosso_jstatmech_05_gaussian}, and the second moment or roughness $B(r)\equiv \sigma_2(r) \sim r^{2\zeta}$ is sufficient to fully characterise the scaling, with a single-valued exponent $\zeta_n=\zeta$ $\forall n$. For multiaffine systems, such as out-of-equilibrium, correlated disorder, or strong pinning scenarios \cite{nattermann_prb_90_creep_domainwall,barabasi_pra_92_multifractiality,kolton_prl_05_flat_interface} the full set of higher order scaling exponents $\zeta_n \neq (n/2)\zeta_2$ are necessary to characterise the interface roughening.

\begin{figure*}[h]
    \centering
    \includegraphics[width=\linewidth]{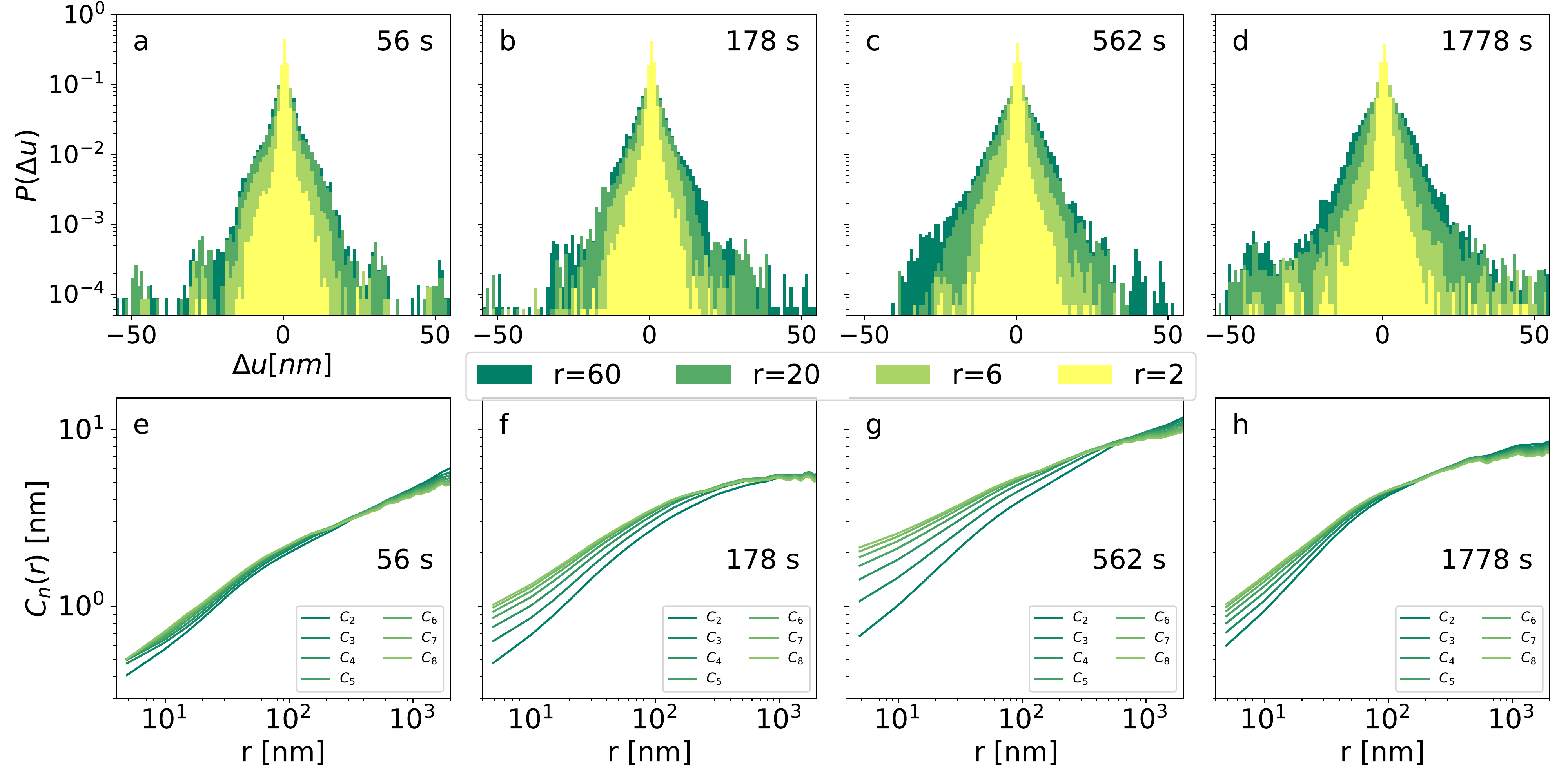}
    \caption{\emph{Multiscaling analysis for domain walls written at 23$^\circ$ C} (a--d) Probability distribution function of relative displacements $\Delta u(r,z)$ for $r \in {6,40,60,400}$ nm and (e--h) power law scaling of its averaged renormalised central moments $C_2$--$C_8$, for domain walls written with switching pulse duration of 56, 178, 562, and 1778 s.}
    \label{PDF_cold}
\end{figure*}

\begin{figure*}[h]
    \centering
    \includegraphics[width=\linewidth]{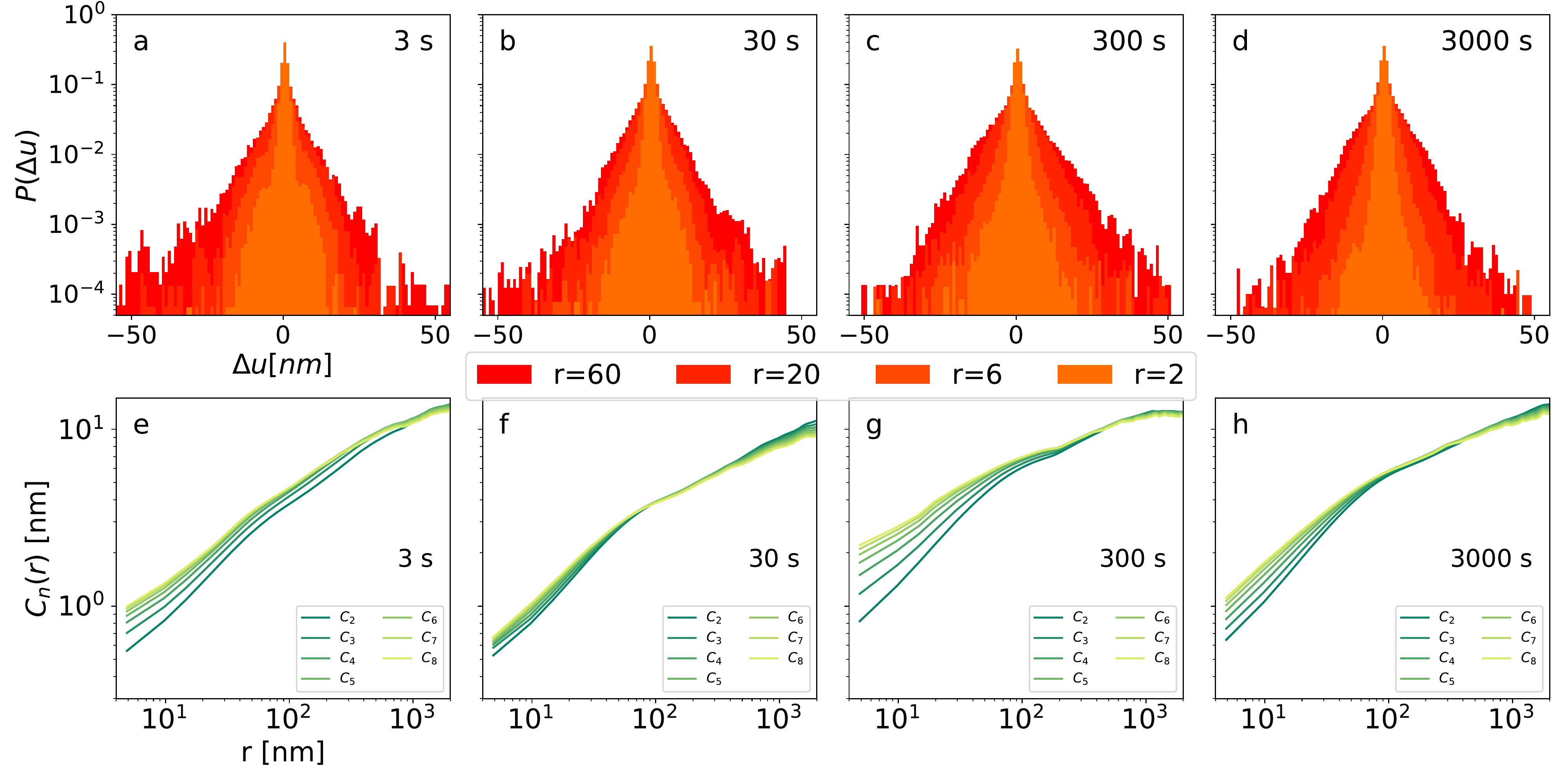}
    \caption{\emph{Multiscaling analysis for domain walls written at 100$^\circ$ C} (a--d) Probability distribution function of relative displacements $\Delta u(r,z)$ for $r \in {6,40,60,400}$ nm and (e--h) power law scaling of its averaged renormalised central moments $C_2$--$C_8$, for domain walls written with switching pulse duration of 3, 30, 300, and 3000 s.}
    \label{PDF_hot}
\end{figure*}

To investigate the nature of the domain walls and the symmetry of their relative displacements, we therefore carried out a multiscaling analysis \cite{santucci_pre_07_fracture_statistics,guyonnet_prl_12_multiscaling} to evaluate their PDF and its central moments. As shown in Figs. \ref{PDF_cold} (a--d) and \ref{PDF_hot}(a--d), the PDFs obtained for different $r$ values and switching pulse durations for domain walls written at 23$^\circ$ C and 100$^\circ$ C, respectively, are generally quite symmetric. The corresponding renormalised central moments, shown in Figs. \ref{PDF_cold} (e--h) and \ref{PDF_hot}(e--h)
\begin{equation}
    C_n(r,z) = [\sigma_n(r,z)]^{1/n}
\end{equation} 
show some fanning at the highest and lowest $r$ values, but collapse to within 15 \% in the intermediate range. 

\end{document}